\documentclass{svmult} 

\usepackage{graphicx}        
\usepackage[bottom]{footmisc}

\font\af=msbm10

\def\R{\hbox{\af R}}
\def\I{\hbox{\af I }}

\begin{document}

\title*{  Path integrals and boundary conditions}
\author{Manuel Asorey \inst{1}\and
Alberto Ibort\inst{2}\and
 Giuseppe Marmo\inst{3}}
\institute{Departamento de F\'\i sica Te\'orica, Facultad de Ciencias, Universidad de
  Zaragoza, 50009 Zaragoza, Spain. 
\texttt{asorey@unizar.es}\and
Depto. de Matem\'aticas, Universidad Carlos III de Madrid.  28911
  Legan\'es, 
Madrid,  Spain. \texttt{albertoi@math.uc3m.es}\and
Dipartamento di Scienze Fisiche, INFN Sezione di Napoli, Univ. di
  Napoli Federico II, 80126 Napoli, Italy. 
\texttt{marmo@na.infn.it}
}

\maketitle

The path integral approach to quantum mechanics provides a method
of quantization of  dynamical systems directly from the Lagrange formalism. 
In field theory the method presents  some advantages over  Hamiltonian 
quantization. The Lagrange formalism  preserves relativistic  covariance 
which makes the Feynman method  very convenient
to achieve the  renormalization of field theories both in 
perturbative and  non-perturbative approaches. However,  when the  
systems are confined in bounded domains we shall show that the path 
integral approach  does not  describe the most general type of
boundary conditions. Highly non-local boundary conditions cannot be
described by  Feynman's approach. We analyse in this note the origin of 
this problem in
quantum mechanics and its implications for field theory.  


\section{Introduction}
\label{sec:1}
The original Heisenberg formulation of quantum mechanics has been
complemented with two equivalent pictures: Schr\"odinger wave mechanics
and Feynman's path integral. Dirac proved that the Heisenberg and Schr\"odinger pictures
both  based on the Hamiltonian approach were equivalent. Feynman's 
path integral
method  although based on the Lagrangian formalism  is also equivalent to the other
two formulations for most quantum mechanical systems \cite{Feynman, Feynman-Higgs}. In field theory Feynman's
formulation proved to be very useful. In perturbation theory, the explicit covariant
character of its natural functional integral generalization makes possible a 
simpler approach to the renormalization of ultraviolet divergences.
 In the Euclidean version \cite{FeyKac} 
the functional integral formulation  is a crucial ingredient for
non-perturbative  approaches to field theory  and critical phenomena
\cite{wilson}.  The equivalence of the functional integral formulation with the
Hamiltonian pictures also holds for constrained systems  like   gauge theories
and string theories. The ordering ambiguity problem of  Hamiltonians
in  constrained systems  has also an analytic counterpart in the functional
method. Indeed,
Ito and Stratonovich discretizations of path integrals provide different
prescriptions for quantum systems. 
However, there is an exception to this rule. For systems constrained  to
bounded domains the Feynman approach does not describe the most general
type of boundary conditions compatible with  Hamiltonian approaches
\cite{sant}. The
analysis of this problem is the main goal of this note. 
 
On the other hand the physics of boundary conditions is
becoming very relevant in quantum gravity, string theory and brane theory.
A large variety  of boundary conditions are
required to describe new physical effects. The
limitations of the scope of functional integral methods might require new
methods to describe these phenomena.
Effects like anomalies \cite{anm, aae, esteve}
topology change \cite{rbal} quantum holography \cite{bek,qg,hol}, quantum gravity 
and AdS/CFT correspondence \cite{mal} 
show the relevance of boundaries in the description
of fundamental physical phenomena. Moreover, the recent observation of a
suppression of quadropole and octopole components of the cosmic
background radiation might be connected with the boundary conditions
or the space topology of the Universe \cite{dod}.
To some extent the role of  boundary phenomena
has been promoted from  academic and phenomelogical simplifications
of  complex physical  systems to a higher status connected with
very basic fundamental principles. Thus, the issue of whether the
path integral approach is able to describe all boundary effects is very important. Otherwise,
 the analysis of non-perturbative effects under generic boundary conditions 
must  rely in  new non-perturbative Hamiltonian methods.




\section{Quantum boundary conditions}
\label{sec:2}


Probability  preservation is the fundamental quantum dynamical principle 
which imposes the more severe constrains on the boundary conditions
of  systems evolving in bounded domains.
The analytical condition, which is encoded by self--adjointness  of the 
Hamiltonian operator, contains all the quantum subtleties associated
to the unitarity principle and the dynamical behaviour at the boundary.


The existence of a boundary generically enhances the genuine quantum aspects 
of the system. Famous  examples of this enhancement are the Young two slits experiments
and the Aharanov-Bohm effect, which pointed out the relevance of  boundary 
conditions  in the quantum theory. Another examples of quantum physical
phenomena which are intimately related to boundary conditions are the Casimir effect
\cite{cas, vac} the role of edge states \cite{sr} and the quantization of conductivity \cite{qcd, 9bis}
in the quantum Hall effect.

Let us consider a point-like particle moving on a bounded domain $\Omega$
of $\R^n$ with regular and oriented boundary $\partial\Omega$. The Schr\"odinger
picture prescribes that the  Hamiltonian is given from the scalar Laplacian $\Delta$. 
This operator is symmetric on the domain of smooth functions with compact support on
$\Omega$. However, in this domain  the operator is not self--adjoint as it is required by the unitarity quantum
principle of    time evolution. Now,  the Hamiltonian operator can be
extended to a larger 
domain of functions where it becomes self--adjoint. 
The extension, however, is not unique.
The  classification of self--adjoint extensions of the Hamiltonian
can be characterized in terms of unitary operators between
 defect subspaces in the classical theory due to von Neumann \cite{ds,gp}.
However, there is a more useful  characterization of these selfadjoint
 extensions in
terms of constraint conditions on the boundary values of the wave functions
\cite{aim}.
 In this framework
the set of {  self--adjoint extensions of the Hamiltonian
 is in one-to-one correspondence with the group of unitary operators of the
 space $L^2(\partial\Omega)$  of square integrable functions of the boundary $\partial\Omega$.}

Thus,  any unitary operator  $U$ on the space of boundary functions  
which are square integrable  with respect to  the standard Riemaniann
measure $d\mu_{_{\partial\Omega}}$ induced from the Euclidean metric of $\R^n$ defines a selfadjoint
of the quantum Hamiltonian $\Delta^U$. Conversely, any selfadjoint extension of
$\Delta$ is associated to one unitary operator of this type \cite{aim}.

The domain of the selfadjoint Hamiltonian governed by $U$ is defined
 by  the wave functions which satisfy the boundary condition
\begin{equation}
 \varphi - i\dot\varphi = U(\varphi + i\dot\varphi)
\label{bc}
\end{equation}
where
$\varphi=\psi|_{\partial\Omega}  $ is the boundary value of the wave function  $\psi$
and $\dot \varphi$ its oriented  normal  derivative at the boundary $\partial\Omega$.
The condition (\ref{bc}) implies the vanishing of the  boundary term remaining
 after integration by parts and use of Stokes theorem which is of the form \cite{aim}
$$\langle \psi_1, \Delta  \psi_2\rangle =
\langle \Delta \psi_1, \psi_2\rangle +i \int_{\partial\Omega} \,
\left[(\dot\varphi_1 , \varphi_2 ) -
(\varphi_1 ,\dot\varphi_2 )\right] d\mu_{_{\partial\Omega}}.
$$
Through  the above characterization, the set of self--adjoint extensions of the
Hamiltonian  inherits  the group structure of the group of unitary operators.
For the half--line the group is $U(1)$ and for a closed interval $U(2)$.
For spaces  of dimension higher than one the group of boundary conditions  
is an infinite dimensional group. There are two particular subsets of  boundary
conditions which can be very explicitly  expressed in
terms of boundary values.  If the spectrum of $U$ does not contain $1$ and 
$-1$, the boundary conditions (\ref{bc}) reduce to
$${\varphi=A_- \dot \varphi ; \qquad {\rm or }\qquad \dot\varphi=A_+  \varphi}$$
where $A_\pm$ are the hermitian operators defined by
the Cayley transform
$$
A_-= i { \I +  U\over \I-U}\qquad A_+= -i { \I -  U\over \I+U}
$$
Notice that the set of boundary conditions where the Cayley transform
becomes singular include two well known types of 
boundary conditions: Dirichlet ($U = - \I$) and Neumann ($U = \I$)
boundary conditions. But the group of boundary conditions is much larger.
In particular, for two dimensional and higher dimensional systems bounded
to compact domains  the space
of selfadjoint extensions is a infinite-dimensional manifold.  The quantum role of 
boundary conditions is very important for the behaviour of low energy
levels. High energy levels are quite independent of boundary effects. Indeed, 
boundary effects play no role in the ultraviolet  regime, whereas they are
crucial for the infrared. In particular, the existence of edge states is only
possible under certain boundary conditions. A very important result 
concerning edge states is that if the 
unitary operator $U$ characterizing the selfadjoint extension
 has one eigenvalue $-1$ with smooth eigenfunction,
the  selfadjoint extensions associated to  $U_t=U {\rm e}^{i t}$ 
have for small values of $t$ one negative energy level which corresponds to an 
edge state. The energy of this edge state becomes infinite when 
$t\to 0$ \cite{aim}. In this case all the negative bounding energy is provided
by the boundary.

\section{Boundary conditions in path integrals}

The action principle governs the classical and quantum dynamics 
of unconstrained systems. The classical dynamics is given by stationary
trajectories from the variational action principle and the quantum dynamics 
is automatically implemented in the path integral formalism by the weight that
the classical action provides for classical trajectories. However,
for particles evolving  in a bounded domain the variational problem is
not uniquely defined. It is necessary to specify the evolution of the particles
after reaching the  boundary. 
The constraints that appear on the trajectories contributing
to the path integral only depend on the  very nature of the physical 
boundary.

In fact,  the nature of the boundary imposes more severe constraints on the
classical dynamics than to the quantum evolution. This is due to the point-like
nature of the particle which requires that after reaching the boundary the
individual particle has to emerge either back  at  the same boundary 
point  or at a  
different point of the same boundary. The only allowed 
freedom  is where it 
emerges back  and the momentum it emerges with. The emergence 
of the particle at a different point covers  the possibility that the domain
be  folded and glued at the apparent boundary  giving rise to
non-trivial topologies. In summary, the classical boundary
conditions are given by two maps:  an isometry of the boundary
$$\alpha:\partial\Omega\to\partial\Omega $$
{\rm and}
a positive density function
$$ \rho:\partial\Omega \to {\rm \R^+}$$
which specify the change of position and normal component of 
momentum of the trajectory of the particle upon reaching the boundary. 
The isometry $\alpha$ encodes the possible geometry and
topology generated by the folding 
of the boundary and the function $\rho$  is associated to 
the reflectivity (transparency or stickiness) properties of the boundary.
Once these two functions are specified the classical variational problem is
restricted to trajectories which satisfy the boundary conditions:
\begin{equation}
{\bf {\bf x}}(t_+)=\alpha({\bf x}(t_-)),
\label{cero}
\end{equation}
\begin{equation}
{\bf n}({\bf x}(t_+)))\cdot\dot 
{\bf x}(t_+) =-\rho({\bf x}(t_-))\, {\bf n}({\bf x}(t_-))\cdot\dot {\bf x}(t_-) 
\label{uno}
\end{equation}
and
\begin{eqnarray}
\label{dos}
 &&
\alpha_\ast( \dot {\bf x}(t_-) -  [{\bf n}({\bf x}(t_-))\cdot\dot {\bf
  x}(t_-)] \, {\bf n}({\bf x}(t_-))
\nonumber\\
&&\phantom{adadfa}
=\dot {\bf x}(t_+)- [{\bf n}({\bf x}(t_+))\cdot \dot {\bf x}(t_+)]  \, {\bf n}({\bf
   x}(t_+)
\end{eqnarray}
for any $t$ such that  ${\bf  x}(t)\in \partial\Omega$, where ${\bf n}$ denotes the exterior normal derivative
at the boundary  $\partial\Omega$ and $$ {\bf x}(t_\pm)=\lim_{s\to0} {\bf  x}(t\pm s).$$


This definition of classical boundary conditions is motivated by
the standard physical heuristic interpretation of boundary
conditions. Linear momentum is not conserved because it is 
partially or totally absorbed by the boundary\footnote{ One  may assume 
that a part of the mass  of the system remains attached at the boundary in 
order to keep energy conservation law. The fraction of the
mass lost   by these contact
interactions depends on $\rho$ which is the stickiness or transparency
functional factor of the boundary, but after many contacts with the 
boundary the whole mass will disappear thoughtout the boundary invalidating
 this very simple heuristic picture.}. 
The major constraints on the choice of 
boundary conditions arise first from  the very notion
of point-like particle which requires that any trajectory which reaches
the boundary has to emerge as a single trajectory from the same 
boundary. The second requirement concerning the permitted 
changes of linear momentum at the boundary is constrained by its
compatibilitywith the action principle. This establishes that 
classical trajectories 
are determined by the stationary points of the classical action
$$S({\bf x})=\int dt\, g_{ij}{\,\dot x^i(t)}{\dot x^j(t)}.$$
The variational principle yields the celebrated Euler-Lagrange motion  
equations $\ddot{\bf x}(t)=0$ provided that the boundary term
\begin{equation}
\sum_{m=1}^N\biggl[ \delta{\bf x}(t_m^+)\cdot\dot 
{\bf x}(t_m^+)- \delta {\bf x}(t_m^-)\cdot\dot {\bf x}(t_m^-) \biggr]
\end{equation}
vanishes, where the sum is over all points  $t_m$  where the trajectories
reach the boundary. The simpler way of fulfilling this requirement is
by imposing the vanishing of each individual term on the sum.
These conditions reduce to the boundary conditions (\ref{uno})(\ref{dos})
provided that the only
permitted variations are tangent to the boundary. In this case the normal
component of $\delta {\bf x}(t_m)$ vanishes, i.e. the  points of trajectories
which reach the boundary are only allowed to move along the boundary.
This conditions is reminiscent of Dirichlet condition for D-branes in 
string theory. There is no analogue of the
Neumann boundary conditions of strings for point-like particles.

Simple but interesting types of boundary conditions already arise in the
Sturm-Liouville problem, $\Omega=[0,1]$. In such a case
the boundary of the configuration space is a discrete  two-points set, 
$\partial\Omega=\{0,1\}$. E{x}amples of classical boundary conditions in
such a case are

 i)$\phantom{ii}$ Neumann  (total absorption): $\alpha=I$, $\rho(0) = \rho(1)= \infty$.

ii) $\phantom{i}$Dirichlet (total reflection): $\alpha=I$, $\rho(0) = \rho(1) = 1$.

iii) Periodic:  $\alpha(0)=1$, $\alpha(1)=0$, $\rho(0) = \rho(1) = 1$.

iv)  Quasi-periodic:  $\alpha(0)=1$, $\alpha(1)=0$,
$\rho(0) = \rho(1) = \epsilon$.

The quantum implementation of those boundary condition is straightforward 
via the path integral method. The only paths to be considered in the Feynman's
path integration are those that satisfy the classical boundary conditions
\footnote{In the Euclidean approach the restrictions on the
paths for Neumann and Dirichlet boundary 
conditions are interchanged with respect to classical boundary
conditions (i)(ii).} (\ref{cero})--(\ref{dos}). 
In the one--dimensional case of Sturm-Liouville problem the space of
quantum boundary conditions   is a four--dimensional Lie  group $U(2)$,
whereas the space of classical boundary conditions is the union of two 
disconnected two--dimensional manifolds,

\begin{eqnarray}
\label{doscon}
{\cal M}_1 &=&\{\psi\in L^2([0,1]), \dot \varphi(0)=({1-\rho_1}) \varphi(1),\dot \varphi(1)=({1-\rho_0})
\varphi(0)  \} \\
\label{doscond}
{\cal M}_0 &=&\{\psi\in L^2([0,1]),\dot \varphi(0)=({1-\rho_0}) \varphi(0), \dot\varphi(1)=({1-\rho_1}) \varphi(1)  \} 
\end{eqnarray}

Thus, the  Feynman path integral approach
does not  cover the whole  set of boundary conditions.

Some quantum boundary conditions which are not of 
type (\ref{doscon})(\ref{doscond}) can be related to periodic boundary
conditions (iii) with  a singular potential supported at the boundary 
\cite{groscheb}. Indeed, 
the boundary condition associated to the unitary operator 
\begin{equation}
U =\frac{1}{2-ia}
\begin{pmatrix}{{i a }&{2}\cr
{2}&{ia}}
\end{pmatrix}
\label{delta}
\end{equation}
can be thought as a delta function potential in a circle, i.e.
usual periodic boundary conditions
$$U =
\begin{pmatrix}{0&1\cr
{1}&0}
\end{pmatrix}
$$
for the Hamiltonian
$$H=-\frac{d^2}{dx^2} + a\delta(x).$$
In other cases, the boundary conditions can be described by periodic
boundary conditions and non-trivial magnetic fluxes, e.g.
in the {\it anti-diagonal} case
$$U=
\begin{pmatrix}
{0&e^{-i\epsilon}\cr
e^{i\epsilon}&0}
\end{pmatrix}
$$
we have (pseudo)periodic boundary conditions
$$
{{\varphi(1)=e^{i\epsilon}\varphi(0)\qquad
\varphi'(1)=e^{i\epsilon}\varphi'(0)} }$$
%
with two opposite probability fluxes propagating across the
boundary. This condition is in fact a topological boundary
condition which corresponds to  fold the interval
into a circle $S^1$, with a magnetic flux $\epsilon$
crossing through the loop, i.e.
the Hamiltonian
$$H=-\left(\frac{d}{dx}-i\epsilon\right)^2$$
with standard periodic boundary conditions.
The images method also permits   to  use unconstrained 
path integral methods to  describe systems with
non-trivial boundary conditions \cite{groschea}.
However, in this case the use of the path integral is
not as simple as in the Feynman original formulation.

But even with  these tricks path integral methods 
do not describe  all possible types of boundary 
conditions even
in the simple case of Sturm-Liouville problem. One of 
reasons behind the failure of path integral picture is 
the single valued nature of trajectories. Many conditions, e.g. the
boundary condition (\ref{delta}) describe a scattering by a singular
potential sitting on the circle \cite{groscheb}. 
There are two different  types of quantum interactions with the boundary: 
reflection and diffraction. A classical
description of the phenomena without including a potential term will
require the  splitting of the ongoing classical trajectory into two outgoing paths:
one pointing forward and another one backwards. This picture destroys 
the pure  point-like particle approach and leads to multivalued trajectories
which dramatically changes  the simple Feynman's description of path
integrals. Furthermore, there are boundary conditions where one single
trajectory upon reaching the boundary has to be split into an infinite
set  of outgoing trajectories. This behaviour can be explicitly
pointed out by noticing that the quantum evolution of a narrow
wave packet evolves backward after being scattered by the  boundary 
as a quite widespread wave packet emerging from all points of the boundary.

In higher dimensions the mismatch between the spaces of quantum and 
classical boundary conditions is even larger \cite{sant}.
It is obvious that there are many quantum  boundary conditions that
cannot be described by local boundary conditions even with the incorporation
of singular potentials which in this case might require renormalization
\cite{dl,0bis}
One particularly interesting example is provided by a particle moving
on an annulus with two circular boundaries (see Figure 1)
\begin{figure}
\centering
\includegraphics[height=4cm]{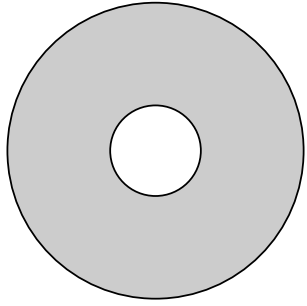}
%
%
\caption{The Corbino disk}
\label{fig:1}       
\end{figure}
with boundary conditions
\begin{equation}
\varphi_\pm=\pm \Delta_{\partial\Omega}
\, \dot\varphi_\pm.
\label{cor}
\end{equation}
 In this case there might appear negative energy levels associated to
edge states \cite{aim}. 
Annular quantum devices (Corbino disks) are  used in some quantum Hall effect 
experiments and the edge states generate chiral currents along the
two edges of the disk.

The boundary condition (\ref{cor}) has no path integral 
description.
In summary, the  Feynman  approach
does not describe the whole  set of boundary conditions.
This fact is a consequence of the enhancement of genuine quantum
effects by the presence of the boundary. The boundary itself
can be considered from this point of view a genuine quantum
device.


\section{Conclusions}
Highly non-local quantum boundary conditions  cannot be described 
by path integrals. This  means that Feynman's approach to quantum
mechanics is not completely equivalent to the Heisenberg and Schr\"odinger
approaches which is a real drawback for the path integral formalism.
One may argue that strictly speaking non-locality might never appears in Nature.
Boundary conditions  for macroscopic boundaries have always a
microscopic origin. But even if the fundamental laws of microscopic 
physics are perfectly local,  effective
non-local macroscopic conditions can be generated in the presence of matter.
The mechanism responsible of the phenomena is  similar 
to what occurs with relativistic invariance.
The microscopic laws of the standard model are
relativistic invariant. However the presence of matter breaks this invariance 
and macroscopic objects like cavities and boundaries do exist. Thus, even if 
the ultimate laws of physics are local, non-local boundary conditions can be
achieved by the presence of highly correlated matter boundaries. In such a case
the quantum system cannot be described by the path integral formalism. 

\section{Acknowledgments}
This article is dedicated to  Alberto Galindo with the occasion of his
retirement.
His deep knowledge of the most intricate aspects of modern physics 
and mathematics and his uncompromising comments and opinions
played a a crucial role in the development of spanish modern 
theoretical physics. The Galindo-Pascual book on quantum mechanics 
\cite{gp} was one of the very few fundamental books where the problem of 
boundary conditions was approached from a rigorous viewpoint. 

%
%
%


\begin{thebibliography}{99.}
%
%
%
\bibitem{Feynman}R. P. Feynman, Rev. Mod. Phys. {\bf 20} (1948) 367
\bibitem{Feynman-Higgs}R. P. Feynman, and  A. R. Hibbs, { \it Quantum mechanics and path integrals}, McGraw-Hill, New York, (1965)
\bibitem{FeyKac}M. Kac, in {\it Proc. 2nd Berkeley Symposium Math. Statist. Probability},
Univ. Cal. Press (1950) 189
\bibitem{wilson}K. Wilson, Phys. Rev. {\bf D10} (1974)2445

\bibitem{sant}M. Asorey,  
in {\it Stochastic processes applied to physics},
 Ed. L. Pesquera and M.A. Rodriguez, 
 World Sci., Singapore (1985)
\bibitem {anm}%
N.S. Manton, Ann. Phys. (N.Y.) {\bf 159} (1985) 220
\bibitem{aae}M. Aguado, M. Asorey and J.G. Esteve,
Commun. Math. Phys. {\bf 218} (2001)
233
\bibitem{esteve} J.G. Esteve, Phys. Rev. {\bf D 34} (1986) 674;
Phys. Rev. {\bf D66}(2002) 125013

\bibitem {rbal}%
 A.P. Balachandran, G. Bimonte,
G. Marmo and A. Simoni, Nucl. Phys. {\bf B 446} (1995) 299
\bibitem{bek} J.
Bekenstein, Lett. Nuovo Cim. {\bf 4} (1972) 737, Phys. Rev. {\bf D7} (1973) 2333
\bibitem {qg}%
G. t' Hooft, in {\it Salamfestschrift: A collection of talks}, 
Eds. A. Ali,J. Ellis and S. 
Randjbar-Daemi World Sci. (1993);   Phys. Scripta {\bf T36} (1991) 247;
Class. Quant.Grav. 16 (1999) 3263
\bibitem {hol}%
L. Susskind, J. Math. Phys. {\bf 36} (1995)6377 
\bibitem {mal}%
J. Maldacena, Adv. Theor. Phys. {\bf 2}(1998) 231
\bibitem {dod}%
J.-P. Luminet, A. Riazuelo, R. Lehoucq and
J.P. Uzan, Nature {\bf 425} (2003) 593
\bibitem {cas}%
H.B.G. Casimir, Proc. K. Ned. Akad. Wet. {\bf 51}(1948) 793
\bibitem {vac}%
P. Milonni, {\it The Quantum Vacuum: An Introduction to Quantum Electrodynamics}
, Academic Press, San Diego (1994)
\bibitem {sr}%
V. John, G. Jungman and S. Vaidya, Nucl. Phys.
{\bf B 455 } (1995) 505

\bibitem {qcd}%
 D.J. Thouless, M. Kohmoto, M.
P. Nightingale and M. den Nijs, Phys. Rev. Lett. {\bf 49} (1982) 
405\bibitem{9bis}
J. Avron and R. Seiler, Phys. Rev. Lett. {\bf 54} (1985)259
\bibitem {ds}%
N. Dunford, J.T. Schwartz, {\it Linear
Operators, Part II: Spectral theory, self--adjoint operators in Hilbert
space}, Wiley, New York (1963)
\bibitem {gp}
A. Galindo and P. Pascual, {\it  Quantum mechanics},
Springer-Verlag, Berlin, 1990-1991
\bibitem{aim}
M. Asorey, A. Ibort and G. Marmo, Int. J. Mod. Phys. {\bf A 12}(2004)
\bibitem{groscheb}C. Grosche,  Phys. Rev. Lett. {\bf 71} (1993) 1

\bibitem{groschea}C. Grosche, Annalen Phys.{\bf 2} (1993) 557 
\bibitem {dl}%
R. Jackiw, in {\it
M.A.B. B\'eg Memorial Volume}, eds. A. Ali and P. Hoodbhoy,
World Sci., Singapore (1992) \bibitem{0bis}
C. Manuel and R. Tarrach, Phys. Lett. {\bf B 301} (1993) 72 












\end{thebibliography}
\end{document}